\documentstyle[12pt,axodraw]{article}
\newcommand{\NP}{Nucl. Phys. }
\newcommand{\PR}{Phys. Rev. }
\newcommand{\PRL}{Phys. Rev. Lett. }
\newcommand{\PL}{Phys. Lett. }

\begin{document}
\baselineskip=20pt

\pagenumbering{arabic}

\begin{flushright}
AS-ITP-2001-001\\
%TU-HEP-TH-??/??
\end{flushright}

\vspace{1.0cm}
\begin{center}
{\Large\sf $\gamma 3\pi$ and $\pi 2\gamma$ form factors from
dynamical constituent quarks}\\[10pt]
\vspace{.5 cm}

{Xiaoyuan Li$^a$, Yi Liao$^{b,c}$}
\vspace{1.0ex}

{\small $a$ Institute of Theoretical Physics, The Chinese
Academy of Sciences,\\Beijing 100080, P.R.China\\}

\vspace{1.0ex}

{\small $b$ Institut f\"ur Theoretische Physik, Universit\"at Leipzig,\\
Augustusplatz 10/11, D-04109 Leipzig, Germany\footnote{Mailing address}\\}

\vspace{1.0ex}

{\small $c$ Department of Physics, Tsinghua University,
Beijing 100084, P.R.China\\}
\vspace{2.0ex}

{\bf Abstract}
\end{center}
We study the form factors of the low energy 
anomalous $\pi 2\gamma$ and $\gamma 3\pi$ processes in the nonlocal
chiral quark model which incorporates the momentum dependence of 
the dynamical quark mass and realizes correctly the chiral symmetries.
The obtained slope parameter for $\pi 2\gamma$ is in reasonable
agreement with the direct experimental results but smaller than 
the ones invoking vector meson dominance. Our result for the $\gamma 3\pi$
form factor interpolates between the two extremes of theoretical
approaches, with the largest one provided by the vector meson dominance
and the smallest one by the Schwinger-Dyson approach.
But all of them are well below the single data point available so far.
This situation will hopefully be clarified by the experiments at CEBAF
and CERN.

\begin{flushleft}


\end{flushleft}

\baselineskip=24pt
\newpage

The $\pi^0\gamma\gamma$ ($\pi 2\gamma$) and 
$\gamma\pi^+\pi^0\pi^-$ ($\gamma 3\pi$) 
processes are the two simplest
chiral anomaly-driven processes that involve electromagnetic
interactions. A consideration of parity conservation, gauge 
invariance and Lorentz
invariance implies the following structures for their amplitudes,
\begin{equation}
\begin{array}{rcl}
\displaystyle{\cal A}^{\pi 2\gamma}_{\mu\nu}&=&
\epsilon_{\mu\nu\rho\sigma}
k^{\rho}_{1}k^{\sigma}_{2}A^{\pi 2\gamma},\\
\displaystyle{\cal A}^{\gamma 3\pi}_{\mu}&=&
\epsilon_{\mu\nu\rho\sigma}
p^{\nu}_{+}p^{\rho}_{0}p^{\sigma}_{-}A^{\gamma 3\pi}.\\
\end{array}
\end{equation}
Here $k_{1,2}$ denote the outgoing momenta of the two photons
with Lorentz indices $\mu$ and $\nu$, and $p_{+,0,-}$ the
incoming momenta of the three pions, for the two processes respectively. 
The dynamical information is encoded
in the form factors $A^{\pi 2\gamma}$ and $A^{\gamma 3\pi}$ which
are Lorentz invariant functions of the relevant momenta. In the low 
energy and chiral limit, they are completely determined by the chiral
anomaly as summarized in the Wess-Zumino-Witten action to be $\cite{wwz}$
\begin{equation}
\begin{array}{rcl}
\displaystyle A^{\pi 2\gamma}_0&=&
\displaystyle \frac{e^2N_c}{12\pi^2 f_{\pi}},\\
\displaystyle A^{\gamma 3\pi}_0&=&
\displaystyle \frac{eN_c}{12\pi^2 f^3_{\pi}},\\
\end{array}
\end{equation}
where $N_c$ and $f_{\pi}$ are respectively the number of colors and
the pion decay constant. Beyond the limit, their dependence on the
relevant momenta is a reflection of the detailed strong dynamics. 
Since these processes involve only one or a few pions, they may provide
an ideal testing ground for models of strong interactions.

The excellent agreement of $A^{\pi 2\gamma}_0$ with the experimental
value extracted from the on-shell decay of $\pi^0\to\gamma\gamma$ had 
historically constituted one of the first pieces of firm evidence that
quarks carry three colors. When one of the photons is off-shell, the form
factor can be parameterized by a slope parameter in the low energy
region,
\begin{equation}
A^{\pi 2\gamma}/A^{\pi 2\gamma}_0=1+a~x,
\end{equation}
where $x=k^2/m^2_{\pi}$ describes the virtuality of the off-shell 
photon with momentum $k$. 
The slope parameter $a$ has been measured both in the time-like 
region of $k$ using the Dalitz decay $\pi^0\to e^+e^-\gamma$ and in
the space-like region through the $\pi^0$ production in $e^+e^-$
collisions. The direct results from TRIUMF and SINDRUM I in the
first category are respectively $a=0.026\pm 0.054$ $\cite{dalitz}$ and 
$a=0.025\pm 0.014({\rm stat.})\pm 0.026({\rm syst.})$ $\cite{sindrum}$. 
The CELLO group
actually measured the form factor in the large space-like region 
and then extracted the slope parameter by extrapolation using the
vector meson dominance to be $a=0.0326\pm 0.0026$ $\cite{cello}$. 
These results are
consistent with each other within the quoted errors. Concerning the
$\gamma 3\pi$ process the experimental situation is less clear.
There has been so far one measurement $\cite{antipov}$
which seems to favor a larger
value of $A^{\gamma 3\pi}_{0}$ than predicted by the chiral anomaly.
Fortunately this situation will be much improved by the experiments
at CEBAF $\cite{cebaf}$ and CERN $\cite{cern}$
which will measure the form factor 
$A^{\gamma 3\pi}$ in a wider range of kinematics. A more precise
value of $A^{\gamma 3\pi}_{0}$ can then be extracted and the form
factor will be available to distinguish the theoretical results 
based on hadronic models.

The low energy physics of the lowest-lying pseudoscalars may be
described by a chiral Lagrangian which is a tower of terms in 
increasing order of energy expansion. The structures of terms at
each order are completely determined by spontaneously broken chiral 
symmetries while their coefficients are left free. These parameters
may be modelled by properly incorporating the relevant degrees of
freedom in the intermediate-energy region. Of special interest in
this regard are the quark-based models which may have a close 
connection to the underlying QCD dynamics. As is well-known, one
feature of dynamical quarks is their running mass in the 
intermediate-energy region, which should have significant effects
on low energy physics when the quarks are integrated out. This point
has been nicely taken into account by Holdom and colloborators in their
nonlocal constituent quark model 
$\cite{nonlocal}\cite{terning}\cite{holdom92}$. 
Indeed, the coefficients in the
$O(p^4)$ chiral Lagrangian for the lowest-lying pseudoscalars
are expressed in terms of convergent
integrals of the quark dynamical mass and their phenomenological 
values are well reproduced. The model has also been successful in
modelling the low energy hadronic contributions to the running QED
coupling at the $Z$ boson pole $\cite{qed}$, 
and in understanding the quark-hadron
duality $\cite{duality}$
and the electroweak couplings of constituent quarks 
themselves $\cite{li}$. 
In this note we shall examine the other aspect of 
dynamical constituent quarks, namely, their implications on the
anomalous sector of the pseudoscalars, especially the form factors
of the $\gamma 3\pi$ and $\pi 2\gamma$ processes . Since the
Ward-Takahashi identities for flavor symmetries in QCD are built 
into the model of Holdom 
{\it et al}. we expect that the form factors so obtained should be
comparable in quality to the coefficients in the $O(p^4)$ chiral 
Lagrangian derived from the model. Our results will be compared
with those based on other approaches.
The main feature here is that
the effects of dynamical quark mass are included in a simplest 
possible form while at the same time avoiding introducing many
free parameters.

The nonlocal constituent quark model is an effective theory in the
intermediate energy regime. In this model all physics is assumed to 
be described by a chiral invariant action quadratic in quark fields. 
The dynamical quark mass $\Sigma(p)$ is incorporated into the action 
and its momentum-dependent nature leads to nonlocal
interactions among dynamical quarks, Goldstone bosons and external 
gauge fields. 
Let us outline the action involving interactions with the external
photon field $A_{\mu}(x)$ relevant to our discussion $\cite{holdom92}$.
The interested reader should consult Refs. 
$\cite{nonlocal}\cite{terning}\cite{holdom92}$
for a complete account. 
\begin{equation}
\begin{array}{rcl}
S&=&\displaystyle\int d^4x~\bar{\psi}i\gamma_{\mu}D^{\mu}\psi
-\int d^4x\int d^4y~\Sigma(x-y)\bar{\psi}(x)\xi(x)
X(x,y)\xi(y)\psi(y),\\
\end{array}
\end{equation}
where $\psi$ represents the up and down quark fields with dynamical
mass $\Sigma(p)$ whose Fourier transform is the quantity $\Sigma(x)$.
And
\begin{equation}
\begin{array}{rcl}
D_{\mu}&=&\partial_{\mu}-ieQA_{\mu},~~Q={\rm diag}(2/3,-1/3),\\
\xi&=&\exp(-i\pi\gamma_5/f_{\pi}),~~\pi=\pi^aT^a,\\
X(x,y)&=&P\exp(-i\int_x^y\Gamma_{\mu}(z)~dz^{\mu}),\\
\Gamma_{\mu}&=&i/2[\xi(\partial_{\mu}-ieQA_{\mu})\xi^{\dagger}
+\xi^{\dagger}(\partial_{\mu}-ieQA_{\mu})\xi]\\
%~~~~
&=&eQA_{\mu}+
i/(2f_{\pi}^2)(\pi\partial_{\mu}\pi-\partial_{\mu}\pi\pi)+\cdots,\\
\end{array}
\end{equation}
where $\pi^a$ is the pion field, $T^a$ is the isospin matrix with
${\rm Tr}[T^aT^b]=\delta_{ab}/2$, and $P$ stands for path-ordering.
For convenience, we list in the following the 
relevant vertices appearing in our calculation of the
$\gamma 3\pi$ and $\pi 2\gamma$ amplitudes. The QED vertex between
quarks and the photon is modified to be
\begin{equation}
\displaystyle ieQ\left[\gamma_{\mu}-(p+p^{\prime})_{\mu}
R(p,p^{\prime})\right],
\end{equation}
where $p(p^{\prime})$ denotes the incoming (outgoing) momentum of
the incoming (outgoing) quark line (same below), and
\begin{equation}
\displaystyle R(p,k)=\frac{\Sigma(p)-\Sigma(k)}{p^2-k^2}.
\end{equation}
We should mention in passing that the appearance of the $R$ term in the
QED vertex just fits the dynamical quark mass $\Sigma$ appearing in
the quark propagator so that the Ward identity still holds.
The pion interaction with quarks is of a familiar form generalized
from the constant mass case,
\begin{equation}
-f^{-1}_{\pi}\gamma_5T^a[\Sigma(p)+\Sigma(p^{\prime})].
\end{equation}
The model generally
contains nonlinear interactions of pions with quarks and photons
due to the nonlinearly realized chiral symmetry and nonlocality.
But we found that for
the processes considered here only the following interaction involving
two pions and two quarks can contribute at one loop level,
\begin{equation}
\begin{array}{rl}
\displaystyle\frac{i}{2f^2_{\pi}}&\left(\{T^a,T^b\}
[\Sigma(p)+\Sigma(p+k_1)+\Sigma(p+k_2)+\Sigma(p^{\prime})]
\right.\\
\displaystyle &\left.+\chi [T^a,T^b]
[\Sigma(p+k_2)-\Sigma(p+k_1)+(k_1-k_2)\cdot(p+p^{\prime})
R(p,p^{\prime})]\right),\\
\end{array}
\end{equation}
where the two pions carry the isospin indices $a,~b$ and the incoming
momenta $k_1,~k_2$ respectively. The parameter $\chi=0,~1$ 
corresponds to the two versions $\cite{terning}\cite{holdom92}$ of
the model. Since it makes little numerical difference, we shall 
henceforth take $\chi=1$, corresponding to Ref. $\cite{holdom92}$.
As one may easily figure out, only the $\chi$ term can contribute
to Fig. 1b for the $\gamma 3\pi$ vertex while the first term in 
Eq. (9) cannot due to symmetry.

Let us consider the two processes whose Feynman diagrams are depicted
in Fig. 1. Since we are interested in the form
factors in the low energy region, we expand the amplitudes in the external
momenta. The leading terms must be the same as predicted by the 
WZW action and thus universal to all models which
correctly incorporate the chiral anomaly. In other words, they must
be independent of the specific form of $\Sigma(p)$.
This is indeed the case.
For example the leading term in the $\pi 2\gamma$ amplitude is
proportional to the following integral
$$\int^{\infty}_{0}dx\frac{d}{dx}\left(\frac{x}{x+\Sigma^2(\sqrt{x})}
\right)^2$$
which is unity independently of $\Sigma$ as long as $\Sigma$ is finite 
in the Euclidean space. For the $\gamma 3\pi$ process the
leading term is contributed only by Fig. 1a, whose integral can be
simplified as
$$\int^{\infty}_{0}\left(-1+\frac{x}{x+\Sigma^2(\sqrt{x})}\right)
d\left(\frac{x}{x+\Sigma^2(\sqrt{x})}\right)^2$$
which is always $-1/3$ for a finite $\Sigma$ in the Euclidean region.
The subleading terms depend explicitly on the integrals
of $\Sigma$ which are collected using Mathematica. $A^{\pi 2\gamma}$
has been parameterized in Eqn. (3). For the $\gamma 3\pi$ process,
as will become clear later on, we need to expand up to the $O(p^4)$ terms
to display the kinematic variation of the form factor.
Using the Bose symmetry we have
\begin{equation}
\displaystyle\frac{A^{\gamma 3\pi}}{A^{\gamma 3\pi}_{0}}=
1+m^{-2}_{\pi}\sum_{i=1}^2b_iS_i
+m^{-4}_{\pi}\sum_{i=1}^6c_iQ_i,
\end{equation}
where $S$ and $Q$ are symmetrized Lorentz invariants of the momenta
$p_{1,2,3}=p_{+,0,-}$,
\begin{equation}
\begin{array}{l}
S_1=p^2_1+p^2_2+p^2_3,\\
S_2=p_1\cdot p_2+p_2\cdot p_3+p_3\cdot p_1,\\
Q_1=(p^2_1)^2+(p^2_2)^2+(p^2_3)^2,\\
Q_2=p^2_1p^2_2+p^2_2p^2_3+p^2_3p^2_1,\\
Q_3=p^2_1p_1\cdot(p_2+p_3)+p^2_2p_2\cdot(p_3+p_1)
+p^2_3p_3\cdot(p_1+p_2),\\
Q_4=p^2_1p_2\cdot p_3+p^2_2p_3\cdot p_1+p^2_3p_1\cdot p_2,\\
Q_5=p_1\cdot p_2p_2\cdot p_3+p_2\cdot p_3p_3\cdot p_1
+p_3\cdot p_1p_1\cdot p_2,\\
Q_6=(p_1\cdot p_2)^2+(p_2\cdot p_3)^2+(p_3\cdot p_1)^2.
\end{array}
\end{equation}
Note that the explicit factors of $m_{\pi}$ are introduced for 
convenience although $m^{-2}_{\pi}b_i$ and $m^{-4}_{\pi}c_i$ actually
do not depend on $m_{\pi}$. 

The coefficients $a$, $b_i$ and $c_i$ are
lengthy integrals involving the dynamical quark mass, which is in turn
related to $f_{\pi}$ by the Pagels-Stokar formula reproduced in the model
$\cite{nonlocal}\cite{terning}\cite{holdom92}$,
\begin{equation}
\displaystyle f^2_{\pi}=\frac{N_c}{4\pi^2}\int^{\infty}_{0}dx
\frac{x(\Sigma-\frac{1}{2}x\Sigma^{\prime})\Sigma}{(x+\Sigma^2)^2},
\end{equation}
with $\Sigma^{\prime}=\frac{d}{dx}\Sigma$. A very simple 
parameterization for $\Sigma(p)$ in the Euclidean space was suggested
by Holdom {\it et al}., which incorporates the correct high energy
behavior of the dynamical mass up to logarithms,
\begin{equation}
\displaystyle\Sigma(p)=\frac{(A+1)m^3}{p^2+Am^2},
\end{equation}
where $m$ is a typical mass scale of the constituent quark and is
related to the parameter $A$ through the Pagels-Stokar formula. 
Fixing $f_{\pi}=84$ MeV in the chiral limit we therefore have only
one free parameter. Since this simple ansatz is quite successful in 
reproducing phenomenological values of low energy quantities as
mentioned previously, it will be used in our numerical analysis
without further adaption.

Our results for the coefficients $a$, $b_i$ and $c_i$ are presented
in Table 1 as a function of the parameter $A$ in the same range of 
values as used previously, where the mass scale $m$ is of order
$300$ MeV. 
Let us first discuss
the slope parameter for the $\pi 2\gamma$ process. We get a stable
result of $a=0.02$ for the range of $A$ in the table. This is in
reasonable consistency with direct results from the
Dalitz decays, but smaller than the one extracted from the large
space-like region by extrapolation using vector meson dominance. 
The slope parameter has been studied in other approaches. The free 
quark loop $\cite{bramon}$ with a constant constituent mass $m$ predicts 
$a=\frac{m^2_{\pi}}{12m^2}$, which is about $0.014$ for $m=330$ MeV.
In the phenomenological approach of vector meson dominance the
momentum dependence of the amplitude derives from the lowest-lying
vector resonances and thus
$a=\frac{m^2_{\pi}}{m^2_{\rho}}\sim 0.03$.
Chiral perturbation theory is appropriate for dealing with low energy
pion-photon interactions, but it is afflicted in the current case
by the unknown counter-term parameters appearing in the $O(p^6)$ 
anomalous chiral Lagrangian. Assuming they are again saturated 
by vector mesons with a mean mass of 
$m^2_V=(9m^2_{\rho}+m^2_{\omega}+2m^2_{\phi})/12$, the sum of loop
and counter-term contributions gives $a=0.032$ $\cite{bijnens}$.
It is clear that our result is larger than the one in the constant 
quark mass model but smaller than the ones (both theoretical and
experimental) invoking vector meson dominance.

\begin{table}
\caption{Results of the coefficients $a$ (in units of $10^{-2}$),
$b_i$ ($10^{-2}$) and $c_i$ ($10^{-3}$) as a function of the 
parameter $A$. The mass $m$ (in units of MeV) is determined
by the Pagels-Stokar formula.
Ignoring the $\chi$ term in Eqn $(9)$ would change $b_i$ and $c_i$
by less than $10\%$.}
\begin{center}
\begin{tabular}{|c|c|c|c|c|c|c|c|c|c|c|}
\hline
$A$ & $m$ & $a$ & $b_1$ & $b_2$ 
& $c_1$ & $c_2$ & $c_3$ & $c_4$ & $c_5$ & $c_6$\\
\hline
$1$   & $342$ & $1.97$ & $-2.29$ & $-2.32$ & 
$-1.02$ & $-1.83$ & $-2.23$ & $-1.96$ & $-2.52$ & $-1.58$\\
$2$   & $317$ & $1.94$ & $-2.37$ & $-2.32$ & 
$-0.97$ & $-1.77$ & $-1.95$ & $-1.71$ & $-1.89$ & $-1.22$\\
$3$   & $299$ & $1.99$ & $-2.52$ & $-2.44$ & 
$-1.09$ & $-1.97$ & $-2.13$ & $-1.85$ & $-1.97$ & $-1.28$\\
$4$   & $287$ & $2.05$ & $-2.67$ & $-2.59$ & 
$-1.27$ & $-2.26$ & $-2.45$ & $-2.12$ & $-2.25$ & $-1.47$\\
$5$   & $277$ & $2.12$ & $-2.82$ & $-2.73$ &
$-1.47$ & $-2.59$ & $-2.83$ & $-2.43$ & $-2.61$ & $-1.71$\\
\hline
\end{tabular}
\end{center}
\end{table}

For the $\gamma 3\pi$ process one has to examine the kinematic
variation of the form factor to extract information on the 
coefficients $b_i$ and $c_i$. In all of the three experiments
available or approved, the photon and two of the pions, which
we assume to be the first and second ones without loss of 
generality, are on-shell, 
$(p_1+p_2+p_3)^2=0$, $p^2_1=p^2_2=m^2_{\pi}$.
(We take $m_{\pi}$ to be the neutral pion mass below and ignore
the small isopin breaking in mass.)
The experiment at Serpukhov and the one at CERN are of
Primakoff type so that the third pion is also on-shell, 
$p^2_3=m^2_{\pi}$, while the CEBAF experiment is to be done at
a low momentum transfer of $p^2_3\approx -m^2_{\pi}$. Defining the
Mandelstam variables $s=(p_1+p_2)^2$, $t=(p_2+p_3)^2$ and 
$u=(p_3+p_1)^2$, the form 
factor is a function of $s$ and $t$ with other kinematic variables
completely fixed. It is then clear that there is no $s$ or $t$
dependence in the $O(p^2)$ terms of $A^{\gamma 3\pi}$ and this is
the reason why we expand up to $O(p^4)$. 

We plot in Fig. 2 our numerical results of the form factor 
$A^{\gamma 3\pi}/A^{\gamma 3\pi}_{0}$ at $A=1$ as a 
function of $s$ with fixed $t=-m^2_{\pi}$, for the Primakoff
case (panel $(a)$) and the CEBAF case (panel $(b)$)
respectively. Also shown are
the results of other approaches, including 
the free quark loop with a constant constituent mass
$\cite{bistrovic}$, the Schwinger-Dyson approach in the generalized
impulse approximation $\cite{alkofer}$, chiral perturbation theory
with vector meson saturation $\cite{cornet}$, vector meson dominance
$\cite{rudaz}$ and its unitarized version $\cite{holstein}$.
The form factors expanded up to second order in $s$ and $t$ in the
free quark loop and the Schwinger-Dyson approaches can be read off
in the original papers. The chiral perturbation
result augmented with vector meson saturation of counter-terms
is $\cite{cornet}$
\begin{equation}
\begin{array}{rcl}
\displaystyle\frac{A^{\gamma 3\pi}}{A^{\gamma 3\pi}_{0}}&=&
\displaystyle 
1+\frac{1}{2m_{\rho}^2}(s+t+u)+\frac{1}{32\pi^2f_{\pi}^2}
\left\{-\frac{1}{3}(s+t+u)\ln\frac{m_{\pi}^2}{m_{\rho}^2}\right.\\
&&\displaystyle \left.+\frac{5}{9}(s+t+u)+\frac{4m_{\pi}^2}{3}
\left[f(m_{\pi}^2,s)+f(m_{\pi}^2,t)+f(m_{\pi}^2,u)\right]\right\},
\end{array}
\end{equation}
where
\begin{equation}
\begin{array}{rcl}
f(m^2,q^2)&=&\displaystyle \left\{
\begin{array}{l}\displaystyle 
                (1-x)z\ln\frac{z+1}{z-1}-2,{\rm ~for~}x<0,\\
                \displaystyle 
         	(1-x)z~2\arctan\frac{1}{z}-2,{\rm ~for~}0<x<1,\\
                \displaystyle 
		(1-x)z\left[\ln\frac{1+z}{1-z}-i\pi\right]-2,{\rm ~for~}1<x,\\
\end{array}
\right.\\
z&=&\displaystyle \sqrt{\left|1-\frac{1}{x}\right|},\\
x&=&\displaystyle \frac{q^2}{4m^2},\\
\end{array}
\end{equation}
which is different from the one quoted for the Primakoff case in Ref. 
$\cite{holstein}$. The phenomenological approach of vector meson dominance
gives $\cite{rudaz}$
\begin{equation}
\displaystyle\frac{A^{\gamma 3\pi}}{A^{\gamma 3\pi}_{0}}=
-\frac{1}{2}\left[1-\left(\frac{m_{\rho}^2}{m_{\rho}^2-s}+
\frac{m_{\rho}^2}{m_{\rho}^2-t}+\frac{m_{\rho}^2}{m_{\rho}^2-u}
\right)\right],
\end{equation}
which is unitarized to be $\cite{holstein}$
\begin{equation}
\begin{array}{rcl}
\displaystyle
\frac{A^{\gamma 3\pi}}{A^{\gamma 3\pi}_{0}}&=&\displaystyle
-\frac{1}{2}\left[1-\left(\frac{m_{\rho}^2}{m_{\rho}^2-s}+
\frac{m_{\rho}^2}{m_{\rho}^2-t}+\frac{m_{\rho}^2}{m_{\rho}^2-u}
\right)\right]
\frac{(m_{\rho}^2-s)(m_{\rho}^2-t)(m_{\rho}^2-u)}
{m_{\rho}^6~D_1(s)D_1(t)D_1(u)},\\
D_1(q^2)&=&\displaystyle
1-\frac{q^2}{m_{\rho}^2}-\frac{q^2}{96\pi^2f_{\pi}^2}
\ln\frac{m_{\rho}^2}{m_{\pi}^2}-\frac{m_{\pi}^2}{24\pi^2f_{\pi}^2}
f(m_{\pi}^2,q^2).
\end{array}
\end{equation}
Note that the results for chiral perturbation and unitarized
vector meson dominance are actually shown for 
$|A^{\gamma 3\pi}/A^{\gamma 3\pi}_0|$ since the form factor can 
become complex in these cases.

It is clear from Fig. $2$ that 
the Schwinger-Dyson approach always gives the lowest values of 
the form factor while the vector meson dominance (especially its
unitarized version) predicts the 
largest values and the steepest change in the kinematic region 
considered here. It is interesting to notice that in contrast to
the case of the vertex $\pi 2\gamma$ the chiral perturbation 
theory predicts a much lower value of the $\gamma 3\pi$ amplitude 
than the vector meson dominance does. Our results interestingly
interpolate the two extremes and are slightly larger than the one 
using a constant quark mass of $330$ MeV.

We have studied the form factors of the low energy 
anomalous $\pi 2\gamma$ and $\gamma 3\pi$ processes in a simple
quark-based model which incorporates the momentum dependence of 
the dynamical quark mass and realizes correctly the chiral symmetries.
The obtained slope parameter for $\pi 2\gamma$ is in reasonable
agreement with the direct experimental results from TRIUMF and 
SINDRUM but smaller than the ones (both theoretical and
experimental) invoking vector meson dominance. All theoretical
predictions for the $\gamma 3\pi$ form factor are well below 
the single data point available so far. But there are also significant
differences among these theoretical results. This situation will 
hopefully be clarified and distinguished by the experiments at 
CEBAF and CERN.

The work of X. Li was supported in part by the China National 
Science Foundation under grant numbers 19835060 and 19875072
and Y. Liao was supported in part by DESY, Germany.
X. Li is grateful to K. Sibold and the staff members of ITP at 
Universit\"at Leipzig for their hospitality during a visit 
when part of the work was done there.
%by the China National Science Foundation under grant number.

%\baselineskip=20pt

\newpage
\begin{flushleft}
{\Large Figure Captions }
\end{flushleft}
\noindent
Fig. 1 Feynman diagrams for the vertices
$\gamma 3\pi$ ($a$ and $b$) and $\pi 2\gamma$ ($c$).
The solid, dashed and wavy lines stand for the quark, pion
and photon fields respectively.

\noindent
Fig. 2 The form factor $A^{\gamma 3\pi}/A^{\gamma 3\pi}_0$
at $A=1$ (solid curve) as a function of $s/m_{\pi}^2$ 
($m_{\pi}=135$ MeV) for the Primakoff case (panel $(a)$) 
and the CEBAF case (panel $(b)$) respectively. 
Also shown are the results of
the following approaches: the free quark loop with a constant
constituent quark mass of $330$ MeV $\cite{bistrovic}$; 
the Schwinger-Dyson approach $\cite{alkofer}$; chiral perturbation
with vector meson saturation $\cite{cornet}$; vector meson
dominance $\cite{rudaz}$ and its unitarization $\cite{holstein}$.

%%%%%%%%%%%%%%%%%%%%%%%%%%%%%%%%%%%%%%%%%%%%%%%%%%%%%%%%%
%%%%     Drawing Figure 1
\newpage
\begin{center}
\begin{picture}(400,600)(0,0)
\SetOffset(10,80)\SetWidth{1.5}

\SetOffset(0,400)
\CArc(100,50)(40,0,360)
\Photon(60,50)(30,50){3}{4}
\DashLine(140,50)(170,50){3}
\DashLine(134.6,70)(170,70){3}
\DashLine(134.6,30)(170,30){3}
\Text(100,-5)[]{\large $(a)$}
\SetOffset(0,-500)

\SetOffset(200,400)
\CArc(100,50)(40,0,360)
\Photon(60,50)(30,50){3}{4}
\DashLine(134.6,70)(170,90){3.}
\DashLine(134.6,70)(170,50){3.}
\DashLine(134.6,30)(170,30){3.}
\Text(100,-5)[]{\large $(b)$}
\SetOffset(-200,-500)

\SetOffset(0,280)
\CArc(100,50)(40,0,360)
\DashLine(60,50)(30,50){3}
\Photon(134.6,70)(170,70){3}{3.5}
\Photon(134.6,30)(170,30){3}{3.5}
\Text(100,-5)[]{\large $(c)$}
\SetOffset(0,-380)

\SetOffset(200,280)
\Text(100,50)[]{\large Figure $1$}

\end{picture}\\
\end{center}
%%%%%%%%%%%%%%%%%%%%%%%%%%%%%%%%%%%%%%%%%%%%%%%%
%%%%%%%%%     Drawing Figure 2
\newpage
\begin{center}
\begin{picture}(350,550)(0,0)

%Panel (a)
%Primakoff type
\SetOffset(40,300)\SetWidth{1.}
\LinAxis(0,0)(300,0)(6,4,5,0,1.5)
\LinAxis(0,200)(300,200)(6,4,-5,0,1.5)
\LinAxis(0,0)(0,200)(5,10,-5,0,1.5)
\LinAxis(300,0)(300,200)(5,10,5,0,1.5)
\Text(0,-10)[]{$4$}
\Text(100,-10)[]{$8$}
\Text(200,-10)[]{$12$}
\Text(300,-10)[]{$16$}
\Text(140,-25)[]{$s/m^2_{\pi}$}
\Text(-15,0)[]{$0.9$}
\Text(-15,40)[]{$1.0$}
\Text(-15,80)[]{$1.1$}
\Text(-15,120)[]{$1.2$}
\Text(-15,160)[]{$1.3$}
\Text(-15,200)[]{$1.4$}
\Text(-40,100)[]
{$\displaystyle\frac{A^{\gamma 3\pi}}{A^{\gamma 3\pi}_0}$}
\Text(15,185)[l]{$(a)$ Primakoff case:}
\Text(35,170)[l]{$p_1^2=p_2^2=p_3^2=m^2_{\pi}$,}
\Text(35,155)[l]{$(p_1+p_2+p_3)^2=0,$}
\Text(35,140)[l]{$t=-m^2_{\pi}$}
%Dynamical quark:
%For A=1
\Curve{(0.000,84.3136)(12.50,84.7477)(25.00,85.2782)
(37.50,85.9052)(50.00,86.6287)(62.50,87.4486)(75.00,88.3650)
(87.50,89.3778)(100.0,90.4872)(112.5,91.6929)(125.0,92.9952)
(137.5,94.3939)(150.0,95.8890)(162.5,97.4806)(175.0,99.1687)
(187.5,100.953)(200.0,102.834)(212.5,104.811)(225.0,106.885)
(237.5,109.056)(250.0,111.322)(262.5,113.686)(275.0,116.146)
(287.5,118.702)(300.0,121.354)
}
%Quark loop:
\Text(250,95)[]{$\cite{bistrovic}$}
\DashCurve{(0.000,77.0400)(12.50,77.4601)(25.00,77.9736)
(37.50,78.5804)(50.00,79.2806)(62.50,80.0742)(75.00,80.9611)
(87.50,81.9414)(100.0,83.0150)(112.5,84.1820)(125.0,85.4423)
(137.5,86.7961)(150.0,88.2432)(162.5,89.7836)(175.0,91.4174)
(187.5,93.1446)(200.0,94.9651)(212.5,96.8790)(225.0,98.8862)
(237.5,100.986)(250.0,103.180)(262.5,105.468)(275.0,107.848)
(287.5,110.322)(300.0,112.890)
}{15}

%Schwinger-Dyson:
\Text(225,53)[]{$\cite{alkofer}$}
\DashCurve{(0.000,51.6240)(12.50,51.8000)(25.00,52.0152)(37.50,52.2695)
(50.00,52.5629)(62.50,52.8955)(75.00,53.2672)(87.50,53.6780)
(100.0,54.1279)(112.5,54.6169)(125.0,55.1451)(137.5,55.7124)
(150.0,56.3188)(162.5,56.9644)(175.0,57.6490)(187.5,58.3728)
(200.0,59.1357)(212.5,59.9378)(225.0,60.7789)(237.5,61.6592)
(250.0,62.5786)(262.5,63.5372)(275.0,64.5348)(287.5,65.5716)
(300.0,66.6475)
}{12}

%Chiral perturbation:
\Text(225,85)[]{$\cite{cornet}$}
\DashCurve{(0.000,67.5232)(12.50,68.6283)(25.00,69.5976)(37.50,70.4613)
(50.00,71.2409)(62.50,71.9523)(75.00,72.6075)(87.50,73.2158)
(100.0,73.7845)(112.5,74.3195)(125.0,74.8257)(137.5,75.3070)
(150.0,75.7666)(162.5,76.2074)(175.0,76.6317)(187.5,77.0415)
(200.0,77.4386)(212.5,77.8245)(225.0,78.2004)(237.5,78.5676)
(250.0,78.9271)(262.5,79.2798)(275.0,79.6264)(287.5,79.9678)
(300.0,80.3046)
}{10}

%Unitarized:
\Text(250,160)[]{$\cite{holstein}$}
\DashCurve{(0.000,67.8500)(12.50,70.4948)(25.00,73.2959)(37.50,76.2835)
(50.00,79.4816)(62.50,82.9099)(75.00,86.5864)(87.50,90.5274)
(100.0,94.7490)(112.5,99.2672)(125.0,104.098)(137.5,109.259)
(150.0,114.769)(162.5,120.646)(175.0,126.911)(187.5,133.586)
(200.0,140.695)(212.5,148.265)(225.0,156.323)(237.5,164.899)
(250.0,174.027)(262.5,183.744)(275.0,194.088)%(287.5,205.102)
%(300.0,216.832)
}{5}

%Vector meson dominance:
\Text(250,125)[]{$\cite{rudaz}$}
\DashCurve{(0.000,62.0740)(12.50,63.1141)(25.00,64.3922)(37.50,65.9125)
(50.00,67.6802)(62.50,69.7011)(75.00,71.9821)(87.50,74.5312)
(100.0,77.3574)(112.5,80.4707)(125.0,83.8826)(137.5,87.6058)
(150.0,91.6545)(162.5,96.0446)(175.0,100.793)(187.5,105.921)
(200.0,111.450)(212.5,117.403)(225.0,123.809)(237.5,130.696)
(250.0,138.100)(262.5,146.058)(275.0,154.611)(287.5,163.808)
(300.0,173.703)
}{2}

%Panel (b)
%CEBAF type
\SetOffset(40,50)\SetWidth{1.}
\LinAxis(0,0)(300,0)(6,4,5,0,1.5)
\LinAxis(0,200)(300,200)(6,4,-5,0,1.5)
\LinAxis(0,0)(0,200)(5,10,-5,0,1.5)
\LinAxis(300,0)(300,200)(5,10,5,0,1.5)
\Text(0,-10)[]{$4$}
\Text(100,-10)[]{$8$}
\Text(200,-10)[]{$12$}
\Text(300,-10)[]{$16$}
\Text(140,-25)[]{$s/m^2_{\pi}$}
\Text(-15,0)[]{$0.9$}
\Text(-15,40)[]{$1.0$}
\Text(-15,80)[]{$1.1$}
\Text(-15,120)[]{$1.2$}
\Text(-15,160)[]{$1.3$}
\Text(-15,200)[]{$1.4$}
\Text(-40,100)[]
{$\displaystyle\frac{A^{\gamma 3\pi}}{A^{\gamma 3\pi}_0}$}
\Text(15,185)[l]{$(b)$ CEBAF case:}
\Text(35,170)[l]{$p_1^2=p_2^2=-p_3^2=m^2_{\pi}$,}
\Text(35,155)[l]{$(p_1+p_2+p_3)^2=0,$}
\Text(35,140)[l]{$t=-m^2_{\pi}$}
\Text(140,-60)[]{\large Figure $2$}
%Dynamical quark:
%For A=1
\Curve{(0.000,55.7359)(12.50,56.3325)(25.00,57.0257)
(37.50,57.8153)(50.00,58.7013)(62.50,59.6838)(75.00,60.7628)
(87.50,61.9383)(100.0,63.2102)(112.5,64.5785)(125.0,66.0433)
(137.5,67.6046)(150.0,69.2624)(162.5,71.0166)(175.0,72.8673)
(187.5,74.8144)(200.0,76.8580)(212.5,78.9980)(225.0,81.2346)
(237.5,83.5675)(250.0,85.9970)(262.5,88.5229)(275.0,91.1453)
(287.5,93.8641)(300.0,96.6794)
}

%Quark loop:
\Text(250,75)[]{$\cite{bistrovic}$}
\DashCurve{(0.000,53.5737)(12.50,54.1806)(25.00,54.8808)
(37.50,55.6743)(50.00,56.5613)(62.50,57.5415)(75.00,58.6152)
(87.50,59.7822)(100.0,61.0425)(112.5,62.3963)(125.0,63.8433)
(137.5,65.3838)(150.0,67.0176)(162.5,68.7448)(175.0,70.5653)
(187.5,72.4792)(200.0,74.4864)(212.5,76.5870)(225.0,78.7810)
(237.5,81.0683)(250.0,83.4490)(262.5,85.9230)(275.0,88.4904)
(287.5,91.1512)(300.0,93.9053)
}{15}

%Schwinger-Dyson:
\Text(200,36)[]{$\cite{alkofer}$}
\DashCurve{(0.000,34.5213)(12.50,34.7886)(25.00,35.0950)(37.50,35.4405)
(50.00,35.8252)(62.50,36.2490)(75.00,36.7119)(87.50,37.2139)
(100.0,37.7550)(112.5,38.3353)(125.0,38.9547)(137.5,39.6132)
(150.0,40.3108)(162.5,41.0476)(175.0,41.8235)(187.5,42.6385)
(200.0,43.4926)(212.5,44.3859)(225.0,45.3183)(237.5,46.2898)
(250.0,47.3004)(262.5,48.3502)(275.0,49.4391)(287.5,50.5671)
(300.0,51.7342)
}{12}

%Chiral perturbation:
\Text(200,57)[]{$\cite{cornet}$}
\DashCurve{(0.000,51.1253)(12.50,52.3724)(25.00,53.4727)(37.50,54.4580)
(50.00,55.3512)(62.50,56.1692)(75.00,56.9248)(87.50,57.6282)
(100.0,58.2872)(112.5,58.9083)(125.0,59.4965)(137.5,60.0564)
(150.0,60.5915)(162.5,61.1047)(175.0,61.5989)(187.5,62.0761)
(200.0,62.5383)(212.5,62.9872)(225.0,63.4242)(237.5,63.8507)
(250.0,64.2678)(262.5,64.6765)(275.0,65.0777)(287.5,65.4723)
(300.0,65.8609)
}{10}

%Unitarized:
\Text(250,155)[]{$\cite{holstein}$}
\DashCurve{(0.000,54.9966)(12.50,58.1083)(25.00,61.3498)(37.50,64.7537)
(50.00,68.3459)(62.50,72.1482)(75.00,76.1801)(87.50,80.4595)
(100.0,85.0039)(112.5,89.8304)(125.0,94.9566)(137.5,100.400)
(150.0,106.181)(162.5,112.318)(175.0,118.834)(187.5,125.751)
(200.0,133.094)(212.5,140.889)(225.0,149.166)(237.5,157.954)
(250.0,167.289)(262.5,177.207)(275.0,187.747)(287.5,198.952)
%(300.0,210.871)
}{5}

%Vector meson dominance:
\Text(250,120)[]{$\cite{rudaz}$}
\DashCurve{(0.000,50.4906)(12.50,51.8689)(25.00,53.4705)(37.50,55.3006)
(50.00,57.3651)(62.50,59.6705)(75.00,62.2244)(87.50,65.0354)
(100.0,68.1131)(112.5,71.4681)(125.0,75.1122)(137.5,79.0588)
(150.0,83.3225)(162.5,87.9196)(175.0,92.8681)(187.5,98.1881)
(200.0,103.901)(212.5,110.033)(225.0,116.611)(237.5,123.665)
(250.0,131.230)(262.5,139.342)(275.0,148.046)(287.5,157.388)
(300.0,167.423)
}{2}

\end{picture}\\
\end{center}

%%%%%%%%%%%%%%%%%%%%%%%%%%%%%%%%%%%%%%%%%%%%%%%%

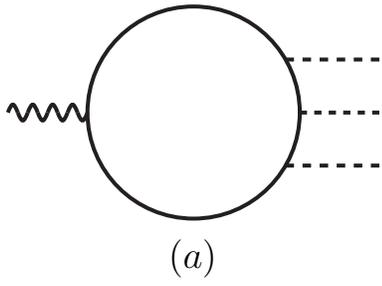
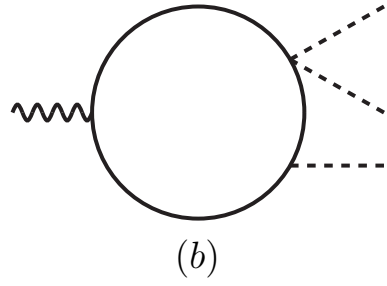
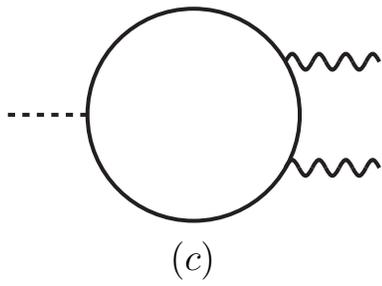
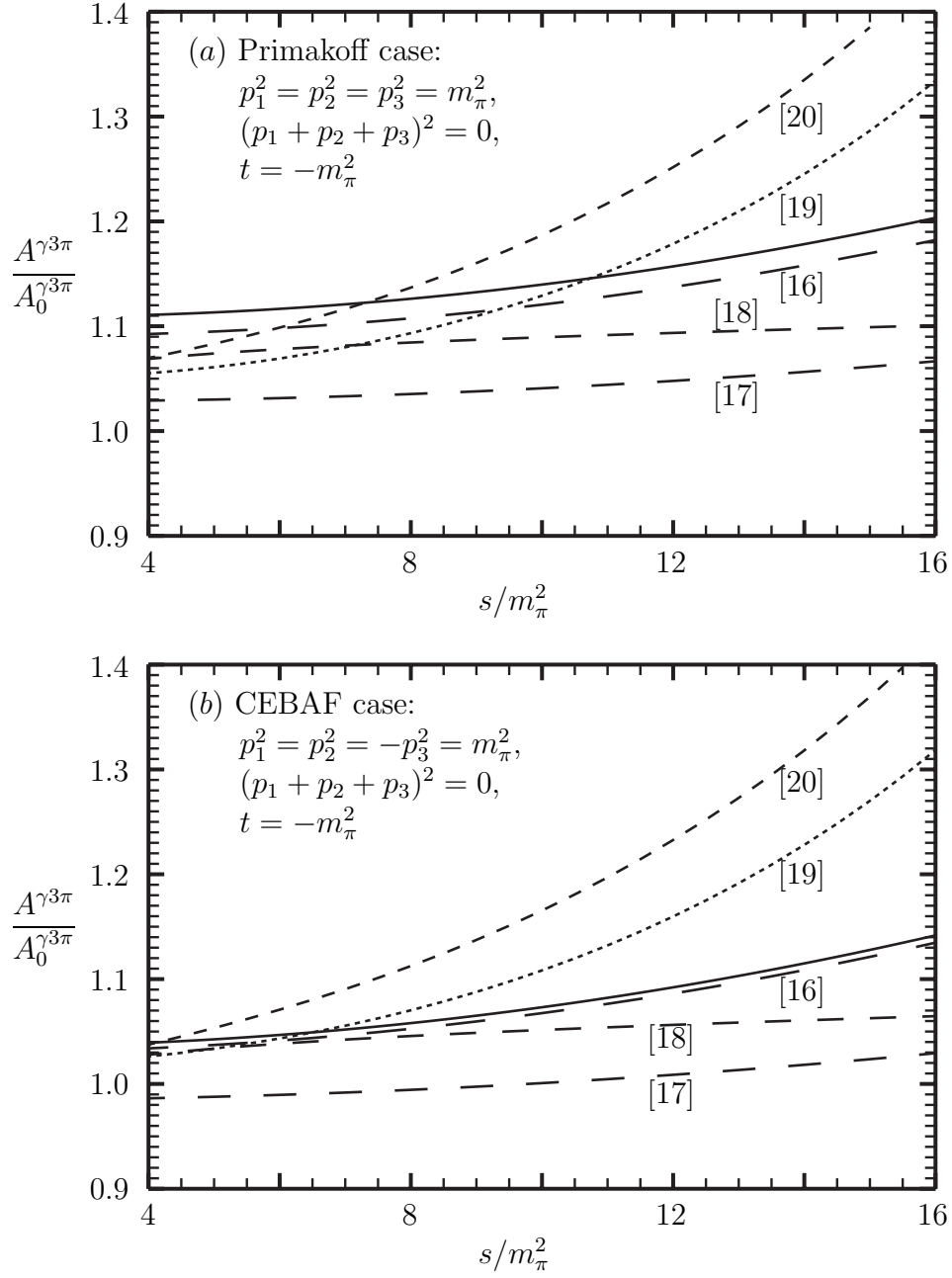
\end{document}